\documentclass[%
 superscriptaddress,
  reprint,
 showpacs,
 showkeys,
 nofootinbib,
  amsmath,amssymb,
 pra,
  longbibliography,
  floatfix,
 ]{revtex4-2}

\usepackage[normalem]{ulem}

\usepackage{adjustbox}

\usepackage{hyperref}
\usepackage{amsmath}
\usepackage{amssymb}
\usepackage{amsthm}
\usepackage{bm} 
\usepackage{graphicx}

\RequirePackage{times}
\RequirePackage{mathptm}

\usepackage{url}
\usepackage[x11names]{xcolor}
\usepackage{eepic}
\usepackage{tikz}
\usetikzlibrary{decorations.markings}
\usetikzlibrary{calc}
\usepackage {pgfplots}
\pgfplotsset {compat=1.8}
\usepackage{epstopdf}
\usepackage[normalem]{ulem}
\theoremstyle{definition}

\begin{document}

\title{Exploitation of Eurosystem Loopholes and Their Quantitative Reconstruction}

\author{Karl Svozil}
\email{karl.svozil@tuwien.ac.at}
\homepage{http://tph.tuwien.ac.at/~svozil}

\affiliation{Institute for Theoretical Physics,
TU Wien,
Wiedner Hauptstrasse 8-10/136,
1040 Vienna,  Austria}

\date{\today}

\begin{abstract}
This paper identifies and analyzes six key strategies used to exploit the Eurosystem's financial mechanisms, and attempts a quantitative reconstruction: inflating TARGET balances, leveraging collateral swaps followed by defaults, diluting self-imposed regulatory rules, issuing money through Emergency Liquidity Assistance (ELA), acquisitions facilitated via the Agreement on Net Financial Assets (ANFA), and the perpetual (re)issuance of sovereign bonds as collateral. The paper argues that these practices stem from systemic vulnerabilities or deliberate opportunism within the Eurosystem. While it does not advocate for illicit activities, the paper highlights significant weaknesses in the current structure and concludes that comprehensive reforms are urgently needed.
\end{abstract}


\maketitle

\section{Introduction}


The inherent intangibility of fiat currencies, such as the Euro, presents opportunities for debasement and resource reallocation strategies.
This issue is exacerbated when nations with diverse economic conditions and demands form a monetary union.
Moreover, the advanced and nuanced nature of these strategies surpasses traditional debasement and counterfeiting methods
typically seen in commodity and (bi)metallic coin systems, such as those experienced by the Latin Monetary Union~\cite{Willis1901,Einaudi2001Jul,flandreau-2000,bussiere-2017},
or tactics like issuing duplicate securities with identical registration numbers and codes.

In the following sections, we will briefly examine each of the six strategies identified,
which could, at least in principle, be implemented by any National Central Bank (NCB) within the Eurosystem---and one potentially even outside it.
Our analysis will delve into detailed considerations only where there is minimal overlap with existing literature.
This includes works addressing topics such as credit facilities among the
Eurosystem's national central banks~\cite{Sinn-Target2,Sinn2017Mar,sinn-2020,blake-2023,perotti-2024,ojeda-2024},
or, from a legislative standpoint, the necessity of a reliable operational framework~\cite{ojeda-2022}.

These strategies include:
Inflating TARGET balances,
exchanging fake collaterals and then defaulting,
diluting self-imposed rules such as debt ceilings and collateral requirements,
money printing through Emergency Liquidity Assistance (ELA),
acquisitions through Agreement on Net Financial Assets (ANFA),
and finally, the perpetual (re)issuing and chaining of sovereign bonds as Eurosystem collateral.

\section{Strategem~1: Inflate TARGET balances}

To quote the European Central Bank's (ECB's) own publication on TARGET imbalances~\cite[Box~4, p.~42]{EuropeanCentralBank2015Sep}:
\textit{``TARGET balances are the net claims and liabilities of the euro area NCBs
vis-\`a-vis the ECB which arise through cross-border payments settled in
central bank money of the respective national banking sectors or the NCBs
themselves and are executed via the common euro area payment platform
known as TARGET.}

\textit{When a bank makes a payment to another bank via TARGET,
the current account of the payer's bank at its NCB is debited and the current account
of the recipient bank at its NCB is credited. If both banks hold their current accounts
at the same NCB there is no net impact on the aggregate account of banks at the
NCB and there are no implications for TARGET balances. However, in the case
of cross-border transactions, the NCB of the paying bank sees a reduction in that
banks account at the NCB, and the NCB of the recipient bank sees an increase
in the recipient bank's account. Such positions are balanced by a TARGET liability
for the first NCB and a TARGET claim for the second NCB. TARGET liabilities and
claims also result from cross-border transactions by NCBs themselves, such as
the purchase or sale of securities held for investment purposes. At the end of each
day, such intra-Eurosystem claims and liabilities are aggregated and netted out
throughout the Eurosystem. This leaves each NCB with a single net bilateral position
vis-\`a-vis the ECB, in the form of a positive or negative TARGET balance. By design,
all the TARGET balances (including the ECB's balance) add up to zero.''
}

Essentially, TARGET imbalances function as an automatic credit facility among the Eurosystem's national central banks (NCBs),
characterized by the following features~\cite{Sinn-Target2,Sinn2017Mar,sinn-2020,blake-2023,perotti-2024,ojeda-2024}:
\begin{enumerate}
\item[(i)] absence of collateral requirements,
\item[(ii)] zero interest rates,
\item[(iii)] no requirement for approval from any authority, including the ECB's Governing Council or the Executive Board,
\item[(iv)] unlimited duration, unlike the US Federal Reserve System's Interdistrict Settlement Accounts, which require a periodic (annual)
settlement\footnote{Interdistrict Settlement Accounts are settled primarily through the transfer of US Treasury securities.
Historically, Gold Certificates were used in Federal Reserve accounting, but they are no longer part of the settlement process today.}, and
\item[(v)] no caps or limits.
\end{enumerate}

Currently, the consequences of a default by one of the participating national banks on this credit facility are undefined.
The ECB's statements do not address the possibility of default, instead implying that imbalances are resolved through mutual, collectivized credit lines among Eurosystem member countries.
This means that, for example, if the Banco de Espa\~na were to default (partially) on this debt, the Bundesbank or the entire Eurosystem would need to absorb the losses, with the extent of the absorption (from all to none) unknown at present.
It is highly probable that significant political pressures will influence the resolution of such a situation.

On request the Austrian OeNB, one of the national banks of the Eurosystem,
issued the following statement (in German, translation by deepl.com)~\cite{2024-09-12-OeNB-email}:
\textit{``There is no list of TARGET balances between individual countries. The ECB acts as the central settlement centre for the central banks participating in TARGET;
at the end of the day, the resulting claims and liabilities between the national central banks are netted out to form a claim on or liability to the ECB.''}
This has been corroborated by a statement of the ECB~\cite{2024-09-18-ECB-email}:
\textit{``Please note that TARGET balances are computed automatically at the end of the day in TARGET Services.
Unfortunately, the data you are enquiring for, i.e. the balance position of each National Central Bank (NCB) vis-\`a-vis all other NCBs, is not available.''}

We therefore conclude that, at the end of each day, the TARGET balances are `collectivized' or netted among all members of the Eurosystem,
and no official record exists to delineate and analyze them, such as a  skew-symmetric matrix with vanishing diagonal entries,
showing the balances of each individual NCB against every other NCB.

This reporting procedure for TARGET balances has significant implications for a potential collapse of the Eurosystem and the subsequent debt settlement
among its member states. The procedure is inadequate and irreversible, as it does not allow for the tracking of individual balances between National Central Banks.
To illustrate this, consider Austria's OeNB, which has a current negative TARGET balance of approximately $-65$ billion (outstanding) Euro~\cite{EuropeanCentralBank2023Jun}.
This balance could have resulted from various (a continuity of)  transactions, such as lending 100 billion to Italy and borrowing 165 billion from Germany,
or alternatively borrowing 30 billion from Italy, lending 100 billion to Germany, and borrowing 135 billion from France.
However, the current system does not allow for the identification of these specific transactions, and no official records are maintained.

Formally, the quantitative flows of inter-NCB Target balances can be represented by a skew-symmetric matrix with zero diagonal elements,
where the entry $T_{ij}$ in the $i$th row and $j$th column represents the respective TARGET balance of the $i$th NCB with respect to the $j$th NCB:
\begin{equation}
T =
\begin{pmatrix}
0 &T_{12}& T_{13} & \cdots & T_{1n}  \\
-T_{12} & 0&T_{23} & \cdots & T_{2n}   \\
-T_{13} & -T_{23} & 0& \cdots & T_{3n} \\
\vdots&\vdots&\vdots&\cdots&\vdots\\
-T_{1n}& -T_{2n}& -T_{3n}&\cdots & 0
\end{pmatrix}
.
\end{equation}
This representation excludes balances of the ECB and the Extra Euro Area (outside of the EU), whose inclusion is straightforward.
There are currently 20 NCBs in the Eurosystem (excluding the ECB), so $n=20$. Including the ECB and the Extra Euro Area would result in $n=21$ and $n=22$, respectively.
The number of independent inter-NCB TARGET balances per day is $(n^2-n)/2= n(n-1)/2$, which equals 190 for $n=20$, 210 for $n=21$, and 231 for $n=22$.

The current TARGET reporting aggregates these balances into $n=20$,
as it only captures the $n$ aggregate sums
\begin{equation}
T_i =
\sum_{j=1}^n T_{ij}
\end{equation}
for the $i$th NCB (noting that $T_{ii}=0$).
All $T_i$ have to sum up to zero, therefore, the parameters $T_i$ are dependent, with $\sum_{i=1}^n T_{i}=0$.
So, there are $n-1$ independent reported aggregate balances, and  $n(n-1)/2$ independent non-reported inter-NCB TARGET balances.
Evidently, since $n(n-1)/2 > n-1$ for $n>2$, this accounting is irreversible for $n>2$, meaning it cannot be reconstructed for more than two
NCBs\footnote{This can be explicitly demonstrated in the case of $n=3$,
as $T_1 = T_{12} + T_{13}$, $T_2 = T_{21} + T_{23} = -T_{12} + T_{23}$, $T_3 = T_{31} + T_{32} = - T_{13} - T_{23}$,
and $T_1 + T_2 + T_3 = 0$
allows for a free (in this scenario unknown inter-NCB TARGET balance) parameter, say $T_{12} \ge 0$, where $T_{13} = -T_{12} - T_2 - T_3$ and $T_{23} = T_{12} + T_2$.}.

\section{Strategem~2: Exchange fake collaterals then default}

Here is a quote from a crisis management paper by the
Financial Stability Institute (FSI) of the
Bank for International Settlements (BIS) exposing this scheme~\cite[p.~9]{Baudino2020Mar}:
\textit{``$\ldots$ banks issued bonds and exchanged them between each other so that
they could be pledged as collateral with the central banks.
This allowed them to break free from central
bank funding limits since they could issue such bonds
(which became known as `love letters') at will.''}

The core strategy behind this approach includes~\cite{Flannery2009,Hreinsson2009,Sibert10}:
\begin{enumerate}
\item[(i)] generating fictitious collateral, such as "I Owe (yo)Us" (IOUs), and
\item[(ii)] subsequently defaulting on these worthless certificates of obligation.
\end{enumerate}

In greater detail, Icelandic banks engaged in a practice where they exchanged debt securities,
using one another's debt as so-called ``love letter'' collateral to secure borrowing from central banks.
The Eurosystem accepted these securities despite concerns about questionable correlation risks.

Between February and April 2008, Icelandic bank subsidiaries borrowed €2.5 billion from the Central Bank of Luxembourg using these ``love letters'' as collateral.
Although the European Central Bank raised concerns and reached an informal agreement to limit the use of such collateral,
the borrowing increased to €4.5 billion by June. By July, the use of ``love letters'' was formally prohibited, leading to a reduction in borrowing to €3.5 billion.

However, by autumn 2008, financial turmoil escalated, and five counterparties defaulted, three of which were Icelandic bank subsidiaries.
Ultimately, in March 2009, following a European Parliament member's inquiry about these loans, the strategem was publicly exposed.

This strategy bears some resemblance to a scheme in which two large European banks outside the Eurosystem were allegedly involved.
During the 2008 financial crisis, these banks reportedly created capital through a credit-share swap.
At the time, they required approximately £7 billion in new capital.

To achieve this, the banks allegedly lent money to a sovereign wealth fund,
which then used the funds to purchase newly issued shares in the banks.
In essence, the banks are said to have lent the sovereign wealth fund the money to reinvest in themselves,
thereby inflating their equity~\cite[t=770]{Werner2016Nov}.
While this approach proved effective in raising capital---as it is unlawful for a bank to lend money to itself~\cite{Westbrook2017Jun}---it was flagged,
albeit unsuccessfully, by one regulatory body.

\section{Strategem~3: Dilute Self-Imposed Rules Such As Debt Ceilings And Collateral Requirements}

The European Central Bank (ECB) has faced increasing scrutiny for diluting its self-imposed rules, particularly regarding debt ceilings and collateral requirements. These rules were initially designed to ensure fiscal responsibility among member states and maintain financial stability in the Eurozone. However, as economic pressures have mounted—especially during crises like the Eurozone debt crisis and the COVID-19 pandemic—the ECB has relaxed these rules to provide more flexibility to member states.

One key area of dilution is the debt-to-GDP ratio.
Originally, Eurozone countries were expected to maintain a debt-to-GDP ratio below 60\%,
as stipulated by the Maastricht Treaty.
However, many countries have consistently exceeded this limit.
Instead of enforcing strict penalties, the ECB has allowed more leniency,
fearing that rigid adherence to these rules could exacerbate economic downturns and stifle growth.

Additionally, the ECB has relaxed collateral requirements,
allowing banks to use riskier assets as collateral for loans.
While this move aims to ensure liquidity in the financial system,
it also raises questions about the quality of assets being accepted.
This leniency has been accompanied by the ECB's practice of indirect monetization of budget deficits,
by buying government debt on secondary markets.
Some fear that by allowing questionable collateral,
the ECB is taking on excessive risk,
which could lead to financial instability if these assets fail to perform.

Here is a quote from an article published by The Brookings Institution~\cite{belz2020ecb}:
\textit{``The ECB now accepts as collateral an expanded set of non-marketable
assets---including government guaranteed loans,
lower quality loans, and small business loans---that are outside of the general framework.
It also granted waivers to Greek sovereign debt, which,
because of its non-investment grade status, was not previously considered eligible collateral.
`Fallen angel' bonds, those that have recently lost their investment-grade rating,
are now accepted as collateral as well.
In addition, the ECB reduced haircuts,
the amount of collateral required in excess of the loan amount,
for its lending programs.
In effect, the ECB decided that it is willing to temporarily
increase its risk tolerance so banks can access the ECB's liquidity operations.''}

At the moment, the central banks hold about one third of the long-term sovereign bonds issued.
This makes them vulnerable against all sorts of attacks by speculators, and against the desires of
the political bodies in their respective countries.
If the trend of collectivizing risks and debt continues, this could, in principle,
make it necessary to refinance these central banks and the Eurosystem in general~\cite{sauer-2023}.

\section{Strategem~4: Moneyprinting Through Emergency Liquidity Assistance (ELA)}

Emergency Liquidity Assistance (ELA) loans are crisis loans embedded in the Eurosystem from its inception,
but their details were only made public in 2017~\cite{EuropeanCentralBank2017Jun}.
A national central bank (NCB) can declare a financial emergency, such as during a bank run or a sovereign debt crisis.

Under such circumstances, NCBs are authorized to print unlimited amounts of money to lend to commercial banks within their jurisdiction,
based on their own collateral rules, to ``rescue'' these banks, allegedly at the NCB's own risk.
This mechanism can also be used to delay bankruptcy and may carry other negative side effects~\cite[Chapter~5]{Sinn2017Mar}.

ELA is a significant resource: For instance, at the height of Greece's 2015 crisis,
ELA borrowing by Greek banks reached 71\%
of the country's nominal GDP~\cite{gibson-2020}.

The ECB Governing Council can block such loans with a two-thirds majority vote.
However, if at least one-third supports the loans, they cannot be stopped.
In 2013, over a third of the ECB Governing Council members were from
GIPSIC (Greece, Ireland, Portugal, Spain, Italy, and Cyprus) countries in need of cheap credit,
making it impossible for others to form a blocking coalition.
Until Latvia's entry in 2014,
these countries could theoretically secure unlimited central bank credit without opposition.


\section{Strategem~5: Aquisitions Through Agreement on Net Financial Assets (ANFA)}

Without public knowledge, national central banks within the Eurosystem were permitted to generate central bank money
for the purpose of acquiring assets on their own account. This encompassed a range of investments tied to currency reserves, employee pension funds,
as well as the pension reserves of the national central banks, along with the counterpart items to statutory capital and reserves.
Furthermore, these assets were utilized for general investment objectives and comprised deposits from governments and international institutions~\cite{Hansen2017}.

The ECB has justified ANFA by invoking the principle of subsidiarity,
allowing national central banks to independently conduct transactions related to national task.
This practice, which amounted to €650 billion in ANFA credits in 2011, caused significant tensions within the ECB council.
As a result, a confidential agreement was made to limit such activities.

Daniel Hoffmann's doctoral thesis~\cite{Hoffmann-2005,Hoffmann-2016}, completed at TU Berlin,
exposed these practices by diligently analyzing accounting records,
ultimately leading to the public release of the details~\cite{EuropeanCentralBank2024Sep,EuropeanCentralBank2022Dec}.

\section{Strategem~6:  Perpetual (Re)Issuing and Chaining of Sovereign Bonds As Eurosystem Collateral}

Sovereign states attempt to chain sovereign bonds~\cite[3]{ECBMemorandum2019Oct},
thereby rolling over increasing amounts of debt without the need for repayment.
In this process, the original principal is devalued by compounded inflation---an effect welcomed by the sovereign debtors---while,
due to compound interest, the required money grows exponentially.

Instead of going into too much detail here I just recall as anecdote the answer of a prominent OECD researcher
to my question
\textit{``will any government ever pay back this sovereign debt, or are they at least committed to do so?''}
at a scientific meeting organized by the Austrian National Bank (OeNB):
his answer was a staightforward \textit{``no''}, after he checked that I am not a journalist.
In that way the sovereign debt can be perpetually rolled over,
and, without disruption, the required amount of money created grows exponentially.

Instead of delving into detail, I mention an anecdote: in response to my question posed to a prominent OECD researcher
at a scientific meeting organized by the Austrian National Bank (OeNB):
\textit{``Will any government ever pay back this sovereign debt, or are they at least committed to doing so?''}
His answer, after affirming I was not a journalist, was a straightforward \textit{``no''}.

This response suggests that sovereign debt can be perceived as sustainable through perpetual rollover, enabling its exponential growth.
The question of whether this ultimately leads to monetary and economic disruptions is an intriguing one~\cite{sauer-2023}.

\section{Reflections of the origin and avoidance of the strategems}

Many of the aforementioned strategies involve securing `free' loans from the Eurosystem, followed by de facto default or postponement of such a default.
With infinite time and volume horizons, repaying the collateral and compounded interest becomes both illusory and irrelevant.

Moreover, if the interest rate is effectively zero, the constraints on purchasing, such as acquiring equity,
are primarily determined by the capacity to obtain credit and the (im)possibility of concealing such transactions from public scrutiny.
 In the (unrealistic) limit, it would theoretically be possible to acquire, and sustain the acquisition of, `everything for nothing' with very little collateral.

The viability of these strategies relies on the use of non-physical fiat currencies,
in contrast to those backed by a limited medium of exchange, such as gold, silver, or algorithmically secured `hard' currencies.
I do not aim to dispute the advantages of fiat money in general,
as I believe that only fiat currency can provide the necessary adaptability and elasticity to cope with productivity increases driven by emerging technologies,
growing populations, and expanding economies.
However, the intangible nature of fiat currencies makes them vulnerable to manipulation,
resulting in unintended wealth redistribution and concentration, as exemplified by the Free Rider Problem~\cite{Congdon2022Mar}.
Consequently, it has been proposed that ``a new constitutional consensus for the EU, endowed with sound economic foundations, is [[$\ldots$]] indispensable''~\cite{ojeda-2022}.

Even if money creation occurs in large quantities, it does not necessarily lead to `excessive' inflation (more than around 100\%
per annum), regardless of the volume of the money supply. While demand-pull inflation---too much money chasing too few goods---might suggest otherwise~\cite{barth-1975},
the possibility of `hoarding'---money stashed away and suddenly released---must also be considered.

Indeed, it has been proposed that as long as this additional money does not appear in foreign exchanges,
competing against other currencies, it will not lead to substantial inflation~\cite{sauer-2023}.
This is exemplified by Japan, which has experienced both large trade surpluses and internal debt~\cite{Werner2003May,Werner2005Mar}.

It is also conceivable that as long as the currency is `demanded' or `wanted' in relation to other currencies---such as due to a trade surplus or demand for commodities,
particularly those related to energy---there will be no excessive inflation.
Consequently, if one holds a reserve currency (which is always `wanted' because of perpetual demand for those resources),
it is possible to print `as much money as one wants' without causing excessive inflation.

The causes and true nature of these scenarios or strategems remain ambiguous to external observers:
are they the product of inadequate design and unforeseen repercussions, or are they deliberately coordinated by factions within or outside the Eurosystem
to, for instance, reassign wealth among member states~\cite{Congdon2022Mar}?
Regardless of whether the impetus stems from systemic deficiencies or calculated maneuvers, the Eurosystem demands~\cite{ojeda-2022} a comprehensive overhaul.

\section{Conclusions}

This paper has identified and analyzed six key strategies employed to exploit the Eurosystem's financial mechanisms,
highlighting significant vulnerabilities within the current structure.
These strategies include inflating TARGET balances,
leveraging collateral swaps followed by defaults,
diluting self-imposed regulatory rules,
issuing money through Emergency Liquidity Assistance (ELA),
acquisitions facilitated via the Agreement on Net Financial Assets (ANFA),
and the perpetual (re)issuance of sovereign bonds as collateral that, through ever increasing compound interest, may enter an unsustainable regime.

The analysis highlights that these practices arise from either unintended systemic vulnerabilities or deliberate opportunism within the Eurosystem.
The intangible nature of fiat currencies poses distinct challenges and opportunities for debasement and resource reallocation,
which are amplified in a monetary union characterized by diverse economic conditions.
This emphasizes the necessity for comprehensive reforms to address these vulnerabilities.

The Eurosystem's current reporting procedures and lack of transparency in tracking individual balances between National Central Banks (NCBs) pose significant challenges.
The irreversible nature of the current TARGET balance reporting makes it impossible to reconstruct detailed inter-NCB transactions,
which is crucial for understanding and mitigating potential defaults and their implications.

Furthermore, the erosion of self-imposed regulations, including debt ceilings and collateral requirements, as well as the utilization of ELA and ANFA,
emphasizes the necessity for more stringent regulatory structures and improved governance.
The continuous (re)issuance of sovereign bonds as collateral also highlights the systemic risks linked to uncontrolled debt accumulation.

In conclusion, the Eurosystem's vulnerabilities necessitate a comprehensive overhaul to ensure financial stability and prevent opportunistic exploitation.
Only through robust reforms can the Eurosystem address these challenges and ensure a more resilient and transparent financial framework for all of its member states.

\subsection*{Acknowledgements}

I gratefully acknowledge invitations to previous ``Volkswirtschaftliche Tagungen'' of the OeNB, the Austrian NCB of the Eurosystem, where some of these issues have been discussed.

The author declares no conflict of interest.

The AI assistants
ChatGPT~4o from OpenAI,
mistral chat of mistral.ai,
as well as Claude from Anthropic were used for grammar and syntax checks.

\bibliography{svozil}

\begin{thebibliography}{35}%
\makeatletter
\providecommand \@ifxundefined [1]{%
 \@ifx{#1\undefined}
}%
\providecommand \@ifnum [1]{%
 \ifnum #1\expandafter \@firstoftwo
 \else \expandafter \@secondoftwo
 \fi
}%
\providecommand \@ifx [1]{%
 \ifx #1\expandafter \@firstoftwo
 \else \expandafter \@secondoftwo
 \fi
}%
\providecommand \natexlab [1]{#1}%
\providecommand \enquote  [1]{``#1''}%
\providecommand \bibnamefont  [1]{#1}%
\providecommand \bibfnamefont [1]{#1}%
\providecommand \citenamefont [1]{#1}%
\providecommand \href@noop [0]{\@secondoftwo}%
\providecommand \href [0]{\begingroup \@sanitize@url \@href}%
\providecommand \@href[1]{\@@startlink{#1}\@@href}%
\providecommand \@@href[1]{\endgroup#1\@@endlink}%
\providecommand \@sanitize@url [0]{\catcode `\\12\catcode `\$12\catcode
  `\&12\catcode `\#12\catcode `\^12\catcode `\_12\catcode `\%12\relax}%
\providecommand \@@startlink[1]{}%
\providecommand \@@endlink[0]{}%
\providecommand \url  [0]{\begingroup\@sanitize@url \@url }%
\providecommand \@url [1]{\endgroup\@href {#1}{\urlprefix }}%
\providecommand \urlprefix  [0]{URL }%
\providecommand \Eprint [0]{\href }%
\providecommand \doibase [0]{https://doi.org/}%
\providecommand \selectlanguage [0]{\@gobble}%
\providecommand \bibinfo  [0]{\@secondoftwo}%
\providecommand \bibfield  [0]{\@secondoftwo}%
\providecommand \translation [1]{[#1]}%
\providecommand \BibitemOpen [0]{}%
\providecommand \bibitemStop [0]{}%
\providecommand \bibitemNoStop [0]{.\EOS\space}%
\providecommand \EOS [0]{\spacefactor3000\relax}%
\providecommand \BibitemShut  [1]{\csname bibitem#1\endcsname}%
\let\auto@bib@innerbib\@empty
\bibitem [{\citenamefont {Willis}(1901)}]{Willis1901}%
  \BibitemOpen
  \bibfield  {author} {\bibinfo {author} {\bibfnamefont {H.~P.}\ \bibnamefont
  {Willis}},\ }\href {https://archive.org/details/historyoflatinmo00willuoft}
  {\emph {\bibinfo {title} {A History of the {L}atin {M}onetary {U}nion: {A}
  Study of International Monetary Action}}}\ (\bibinfo  {publisher} {University
  of Chicago Press},\ \bibinfo {address} {Chicago, IL, USA},\ \bibinfo {year}
  {1901})\BibitemShut {NoStop}%
\bibitem [{\citenamefont {Einaudi}(2001)}]{Einaudi2001Jul}%
  \BibitemOpen
  \bibfield  {author} {\bibinfo {author} {\bibfnamefont {L.}~\bibnamefont
  {Einaudi}},\ }\href {https://doi.org/10.1093/oso/9780199243662.001.0001}
  {\emph {\bibinfo {title} {Money and Politics: {E}uropean Monetary Unification
  and the International Gold Standard (1865-1873)}}}\ (\bibinfo  {publisher}
  {Oxford University Press},\ \bibinfo {address} {Oxford, England, UK},\
  \bibinfo {year} {2001})\BibitemShut {NoStop}%
\bibitem [{\citenamefont {Flandreau}(2000)}]{flandreau-2000}%
  \BibitemOpen
  \bibfield  {author} {\bibinfo {author} {\bibfnamefont {M.}~\bibnamefont
  {Flandreau}},\ }\bibfield  {title} {\bibinfo {title} {The economics and
  politics of monetary unions: {A} reassessment of the {L}atin {M}onetary
  {U}nion, 1865-71},\ }\href {https://doi.org/10.1017/s0968565000000020}
  {\bibfield  {journal} {\bibinfo  {journal} {Financial History Review}\
  }\textbf {\bibinfo {volume} {7}},\ \bibinfo {pages} {25} (\bibinfo {year}
  {2000})}\BibitemShut {NoStop}%
\bibitem [{\citenamefont {Preda}(2017)}]{bussiere-2017}%
  \BibitemOpen
  \bibfield  {author} {\bibinfo {author} {\bibfnamefont {D.}~\bibnamefont
  {Preda}},\ }\href {https://doi.org/10.3726/b10858} {\emph {\bibinfo {title}
  {The history of the {E}uropean {M}onetary {U}nion}}},\ \bibinfo {series}
  {Euroclio}, Vol.~\bibinfo {volume} {99}\ (\bibinfo  {publisher} {P.I.E. Peter
  Lang},\ \bibinfo {address} {Bruxelles, Bern, Berlin, Frankfurt am Main, New
  York, Oxford, Wien, EU},\ \bibinfo {year} {2017})\BibitemShut {NoStop}%
\bibitem [{\citenamefont {Sinn}\ and\ \citenamefont
  {Wollmersh\"auser}(2012)}]{Sinn-Target2}%
  \BibitemOpen
  \bibfield  {author} {\bibinfo {author} {\bibfnamefont {H.-W.}\ \bibnamefont
  {Sinn}}\ and\ \bibinfo {author} {\bibfnamefont {T.}~\bibnamefont
  {Wollmersh\"auser}},\ }\bibfield  {title} {\bibinfo {title} {Target loans,
  current account balances and capital flows: {T}he {ECB}'s rescue facility},\
  }\href {https://doi.org/10.1007/s10797-012-9236-x} {\bibfield  {journal}
  {\bibinfo  {journal} {International Tax and Public Finance}\ }\textbf
  {\bibinfo {volume} {19}},\ \bibinfo {pages} {468} (\bibinfo {year}
  {2012})}\BibitemShut {NoStop}%
\bibitem [{\citenamefont {Sinn}(2017)}]{Sinn2017Mar}%
  \BibitemOpen
  \bibfield  {author} {\bibinfo {author} {\bibfnamefont {H.-W.}\ \bibnamefont
  {Sinn}},\ }\href {https://doi.org/10.1093/acprof:oso/9780198702139.001.0001}
  {\emph {\bibinfo {title} {{The Euro Trap: On Bursting Bubbles, Budgets, and
  Beliefs}}}}\ (\bibinfo  {publisher} {Oxford University Press},\ \bibinfo
  {address} {Oxford, England, UK},\ \bibinfo {year} {2017})\BibitemShut
  {NoStop}%
\bibitem [{\citenamefont {Sinn}(2020)}]{sinn-2020}%
  \BibitemOpen
  \bibfield  {author} {\bibinfo {author} {\bibfnamefont {H.-W.}\ \bibnamefont
  {Sinn}},\ }\href {https://doi.org/10.1007/978-3-030-50170-9} {\emph {\bibinfo
  {title} {The economics of {TARGET} balances: {F}rom {L}ehman to {C}orona}}}\
  (\bibinfo  {publisher} {Palgrave Macmillan imprint published by the
  registered company Springer Nature Switzerland AG},\ \bibinfo {address}
  {Cham, Switzerland},\ \bibinfo {year} {2020})\BibitemShut {NoStop}%
\bibitem [{\citenamefont {Blake}(2023)}]{blake-2023}%
  \BibitemOpen
  \bibfield  {author} {\bibinfo {author} {\bibfnamefont {D.}~\bibnamefont
  {Blake}},\ }\bibfield  {title} {\bibinfo {title} {{Target2}: {T}he silent
  bailout system that keeps the {E}uro afloat},\ }\href
  {https://doi.org/10.3390/jrfm16120506} {\bibfield  {journal} {\bibinfo
  {journal} {Journal of Risk and Financial Management}\ }\textbf {\bibinfo
  {volume} {16}},\ \bibinfo {pages} {506} (\bibinfo {year} {2023})}\BibitemShut
  {NoStop}%
\bibitem [{\citenamefont {Perotti}(2024)}]{perotti-2024}%
  \BibitemOpen
  \bibfield  {author} {\bibinfo {author} {\bibfnamefont {R.}~\bibnamefont
  {Perotti}},\ }\bibfield  {title} {\bibinfo {title} {Understanding the
  {G}erman criticism of {T}arget},\ }\href
  {https://doi.org/10.1093/epolic/eiae009} {\bibfield  {journal} {\bibinfo
  {journal} {Economic Policy}\ }\textbf {\bibinfo {volume} {38}},\ \bibinfo
  {pages} {827} (\bibinfo {year} {2024})}\BibitemShut {NoStop}%
\bibitem [{\citenamefont {Ojeda}(2024)}]{ojeda-2024}%
  \BibitemOpen
  \bibfield  {author} {\bibinfo {author} {\bibfnamefont {A.~R.}\ \bibnamefont
  {Ojeda}},\ }\bibfield  {title} {\bibinfo {title} {El {E}urosistema:
  dominancia monetaria o redistribuci\'on mediante regulaci\'on. {E}n especial,
  los saldos {TARGET2} ({T}he {E}urosystem: monetary dominance or
  redistribution through regulation. {I}n particular, {TARGET2} balances)},\
  }\href {https://doi.org/10.31009/indret.2024.i4.11} {\bibfield  {journal}
  {\bibinfo  {journal} {InDret (Revista para el An\'alisis del Derecho)}\
  }\textbf {\bibinfo {volume} {4}},\ \bibinfo {pages} {349} (\bibinfo {year}
  {2024})}\BibitemShut {NoStop}%
\bibitem [{\citenamefont {Ojeda}(2022)}]{ojeda-2022}%
  \BibitemOpen
  \bibfield  {author} {\bibinfo {author} {\bibfnamefont {A.~R.}\ \bibnamefont
  {Ojeda}},\ }\bibfield  {title} {\bibinfo {title} {Which constitutional
  economics for the two-decade-old eurozone?},\ }\href
  {https://feu-journal.eu/wp-content/uploads/2023/03/FEU-Journal-Issue-2\_Inflation-rising.pdf}
  {\bibfield  {journal} {\bibinfo  {journal} {Future Europe Journal}\ }\textbf
  {\bibinfo {volume} {2}},\ \bibinfo {pages} {42} (\bibinfo {year}
  {2022})}\BibitemShut {NoStop}%
\bibitem [{\citenamefont {{European Central
  Bank}}(2015)}]{EuropeanCentralBank2015Sep}%
  \BibitemOpen
  \bibfield  {author} {\bibinfo {author} {\bibnamefont {{European Central
  Bank}}},\ }\href {https://www.ecb.europa.eu/pub/pdf/ecbu/eb201506.en.pdf}
  {\bibinfo {title} {{Economic Bulletin Issue 6 / 2015}}} (\bibinfo {year}
  {2015}),\ \bibinfo {note} {european Central Bank, publishing date September
  17, 2015, accessed September 8th, 2024}\BibitemShut {NoStop}%
\bibitem [{\citenamefont {Ackerler}(2024)}]{2024-09-12-OeNB-email}%
  \BibitemOpen
  \bibfield  {author} {\bibinfo {author} {\bibfnamefont {M.}~\bibnamefont
  {Ackerler}},\ }\href@noop {} {\bibinfo {title} {{OeNB-Website: Matrix der
  Target Salden nach L\"andern (pro Jahr)?}}} (\bibinfo {year} {2024}),\
  \bibinfo {note} {{A}bteilung f\"ur Kommunikation, Oesterreichische
  Nationalbank, Otto-Wagner-Platz 3, 1090 Wien, Austria, EU, email message,
  dated Sept. 12, 2024}\BibitemShut {NoStop}%
\bibitem [{\citenamefont {Putignano}(2024)}]{2024-09-18-ECB-email}%
  \BibitemOpen
  \bibfield  {author} {\bibinfo {author} {\bibfnamefont {A.}~\bibnamefont
  {Putignano}},\ }\href@noop {} {\bibinfo {title} {{accounting of TARGET ({\#}4
  -- 175953)}}} (\bibinfo {year} {2024}),\ \bibinfo {note} {public
  Communication, European Central Bank, email message, dated Sept. 18,
  2024}\BibitemShut {NoStop}%
\bibitem [{\citenamefont {{European Central
  Bank}}(2024{\natexlab{a}})}]{EuropeanCentralBank2023Jun}%
  \BibitemOpen
  \bibfield  {author} {\bibinfo {author} {\bibnamefont {{European Central
  Bank}}},\ }\href
  {https://www.ecb.europa.eu/stats/policy\_and\_exchange\_rates/target\_balances/html/index.en.html}
  {\bibinfo {title} {{TARGET balances of participating NCBs}}} (\bibinfo {year}
  {2024}{\natexlab{a}}),\ \bibinfo {note} {accessed October 21st,
  2024}\BibitemShut {NoStop}%
\bibitem [{\citenamefont {Baudino}\ \emph {et~al.}(2020)\citenamefont
  {Baudino}, \citenamefont {Sturluson},\ and\ \citenamefont
  {Svoronos}}]{Baudino2020Mar}%
  \BibitemOpen
  \bibfield  {author} {\bibinfo {author} {\bibfnamefont {P.}~\bibnamefont
  {Baudino}}, \bibinfo {author} {\bibfnamefont {J.~T.}\ \bibnamefont
  {Sturluson}},\ and\ \bibinfo {author} {\bibfnamefont {J.-P.}\ \bibnamefont
  {Svoronos}},\ }\href {https://www.bis.org/fsi/fsicms1.htm} {\bibinfo {title}
  {The banking crisis in {I}celand}} (\bibinfo {year} {2020}),\ \bibinfo {note}
  {fSI Crisis Management Series, No 1, March 26, 2020, accessed Sept. 6,
  2024}\BibitemShut {NoStop}%
\bibitem [{\citenamefont {Flannery}(2009)}]{Flannery2009}%
  \BibitemOpen
  \bibfield  {author} {\bibinfo {author} {\bibfnamefont {M.~J.}\ \bibnamefont
  {Flannery}},\ }\bibfield  {title} {\bibinfo {title} {{I}celand's failed
  banks: {A} post-mortem}} (\bibinfo {year} {2009}),\ \bibinfo {note} {report
  prepared for the Icelandic Special Investigation Commission, 9 March
  2009}\BibitemShut {NoStop}%
\bibitem [{\citenamefont {Hreinsson}\ \emph {et~al.}(2009)\citenamefont
  {Hreinsson}, \citenamefont {Tryggvi},\ and\ \citenamefont
  {Sigridur}}]{Hreinsson2009}%
  \BibitemOpen
  \bibfield  {author} {\bibinfo {author} {\bibfnamefont {P.}~\bibnamefont
  {Hreinsson}}, \bibinfo {author} {\bibfnamefont {G.}~\bibnamefont {Tryggvi}},\
  and\ \bibinfo {author} {\bibfnamefont {B.}~\bibnamefont {Sigridur}},\
  }\bibfield  {title} {\bibinfo {title} {Causes of the collapse of the
  icelanldic banks - responsibility, mistakes and negligence}} (\bibinfo {year}
  {2009}),\ \bibinfo {note} {chapter 21, pp. 1-160, of a report prepared for
  the Icelandic Special Investigation Commission (SIC) to Althingi, 9 March
  2009}\BibitemShut {NoStop}%
\bibitem [{\citenamefont {Sibert}(2010)}]{Sibert10}%
  \BibitemOpen
  \bibfield  {author} {\bibinfo {author} {\bibfnamefont {A.}~\bibnamefont
  {Sibert}},\ }\bibfield  {title} {\bibinfo {title} {Love letters from
  {I}celand: Accountability of the {E}urosystem}} (\bibinfo {year} {2010}),\
  \bibinfo {note} {vox posting from 18 May 2010}\BibitemShut {NoStop}%
\bibitem [{\citenamefont {Werner}(2016)}]{Werner2016Nov}%
  \BibitemOpen
  \bibfield  {author} {\bibinfo {author} {\bibfnamefont {R.}~\bibnamefont
  {Werner}},\ }\href {https://youtu.be/MechH0ebs\_c} {\bibinfo {title} {{Prof.
  Richard Werner - Banking Industry Exposed {\&} Solutions Presented - Dublin
  April 2016}}} (\bibinfo {year} {2016}),\ \bibinfo {note} {youTube, Public
  Banking Forum of Ireland, November 28, 2016, accessed September 12,
  2024}\BibitemShut {NoStop}%
\bibitem [{\citenamefont {Westbrook}(2017)}]{Westbrook2017Jun}%
  \BibitemOpen
  \bibfield  {author} {\bibinfo {author} {\bibfnamefont {I.}~\bibnamefont
  {Westbrook}},\ }\href {https://www.bbc.com/news/business-40341611} {\bibinfo
  {title} {Why has {B}arclays been charged?}} (\bibinfo {year} {2017}),\
  \bibinfo {note} {{BBC} News, dated June 20, 2017, accessed September 12,
  2024}\BibitemShut {NoStop}%
\bibitem [{\citenamefont {Belz}\ \emph {et~al.}(2020)\citenamefont {Belz},
  \citenamefont {Cheng}, \citenamefont {Wessel}, \citenamefont {Gros},\ and\
  \citenamefont {Capolongo}}]{belz2020ecb}%
  \BibitemOpen
  \bibfield  {author} {\bibinfo {author} {\bibfnamefont {S.}~\bibnamefont
  {Belz}}, \bibinfo {author} {\bibfnamefont {J.}~\bibnamefont {Cheng}},
  \bibinfo {author} {\bibfnamefont {D.}~\bibnamefont {Wessel}}, \bibinfo
  {author} {\bibfnamefont {D.}~\bibnamefont {Gros}},\ and\ \bibinfo {author}
  {\bibfnamefont {A.}~\bibnamefont {Capolongo}},\ }\href
  {https://www.brookings.edu/articles/whats-the-ecb-doing-in-response-to-the-covid-19-crisis/}
  {\bibinfo {title} {What's the {ECB} doing in response to the {COVID-19}
  crisis?}} (\bibinfo {year} {2020}),\ \bibinfo {note} {the Brookings
  Institution, published June 4, 2020, accessed Sept 6, 2024}\BibitemShut
  {NoStop}%
\bibitem [{\citenamefont {Sauer}(2023)}]{sauer-2023}%
  \BibitemOpen
  \bibfield  {author} {\bibinfo {author} {\bibfnamefont {I.}~\bibnamefont
  {Sauer}},\ }\bibfield  {title} {\bibinfo {title} {{The Lessons from 1923 for
  the Euro Area: Enlightening the Dark Side of (In-) Solvent Central Banks'
  Balance Sheets}},\ }\href {https://doi.org/10.2139/ssrn.4620462} {\bibfield
  {journal} {\bibinfo  {journal} {SSRN Electronic Journal}\ ,\ \bibinfo {pages}
  {1}} (\bibinfo {year} {2023})}\BibitemShut {NoStop}%
\bibitem [{\citenamefont {{European Central
  Bank}}(2017)}]{EuropeanCentralBank2017Jun}%
  \BibitemOpen
  \bibfield  {author} {\bibinfo {author} {\bibnamefont {{European Central
  Bank}}},\ }\href
  {https://www.ecb.europa.eu/pub/pdf/other/Agreement\_on\_emergency\_liquidity\_assistance\_20170517.en.pdf}
  {\bibinfo {title} {{Agreement on emergency liquidity assistance}}} (\bibinfo
  {year} {2017}),\ \bibinfo {note} {european Central Bank, publishing date June
  16, 2017, accessed September 8th, 2024}\BibitemShut {NoStop}%
\bibitem [{\citenamefont {Gibson}\ \emph {et~al.}(2020)\citenamefont {Gibson},
  \citenamefont {Hall}, \citenamefont {Petroulas}, \citenamefont
  {Spiliotopoulos},\ and\ \citenamefont {Tavlas}}]{gibson-2020}%
  \BibitemOpen
  \bibfield  {author} {\bibinfo {author} {\bibfnamefont {H.~D.}\ \bibnamefont
  {Gibson}}, \bibinfo {author} {\bibfnamefont {S.~G.}\ \bibnamefont {Hall}},
  \bibinfo {author} {\bibfnamefont {P.}~\bibnamefont {Petroulas}}, \bibinfo
  {author} {\bibfnamefont {V.}~\bibnamefont {Spiliotopoulos}},\ and\ \bibinfo
  {author} {\bibfnamefont {G.~S.}\ \bibnamefont {Tavlas}},\ }\bibfield  {title}
  {\bibinfo {title} {The effect of emergency liquidity assistance {(ELA)} on
  bank lending during the {E}euro area crisis},\ }\href
  {https://doi.org/10.1016/j.jimonfin.2020.102154} {\bibfield  {journal}
  {\bibinfo  {journal} {Journal of International Money and Finance}\ }\textbf
  {\bibinfo {volume} {108}},\ \bibinfo {pages} {102154} (\bibinfo {year}
  {2020})}\BibitemShut {NoStop}%
\bibitem [{\citenamefont {Hansen}\ and\ \citenamefont
  {Meyer}(2017)}]{Hansen2017}%
  \BibitemOpen
  \bibfield  {author} {\bibinfo {author} {\bibfnamefont {A.}~\bibnamefont
  {Hansen}}\ and\ \bibinfo {author} {\bibfnamefont {D.}~\bibnamefont {Meyer}},\
  }\bibfield  {title} {\bibinfo {title} {{ANFA} {\textendash} {N}ational money
  creation as an existential threat to the currency union?},\ }\href
  {https://www.intereconomics.eu/contents/year/2017/number/4/article/anfa-national-money-creation-as-an-existential-threat-to-the-currency-union.html}
  {\bibfield  {journal} {\bibinfo  {journal} {Intereconomics}\ }\textbf
  {\bibinfo {volume} {2017}},\ \bibinfo {pages} {230} (\bibinfo {year}
  {2017})}\BibitemShut {NoStop}%
\bibitem [{\citenamefont {Hoffmann}(2015)}]{Hoffmann-2005}%
  \BibitemOpen
  \bibfield  {author} {\bibinfo {author} {\bibfnamefont {D.}~\bibnamefont
  {Hoffmann}},\ }\emph {\bibinfo {title} {{D}ie {EZB} in der {K}rise. {E}ine
  {A}nalyse der wesentlichen {S}onderma{\ss}nahmen von 2007 bis 2012}},\
  \href@noop {} {Ph.D. thesis},\ \bibinfo  {school} {TU Berlin}, \bibinfo
  {address} {Berlin, Germany, EU} (\bibinfo {year} {2015}),\ \bibinfo {note}
  {westarp BookOnDemand, Pro Business Verlag, Artikelnummer: 14603, Nr. 1 der
  Schriftenreihe zur Erforschung des Geldwesens}\BibitemShut {NoStop}%
\bibitem [{\citenamefont {Hoffmann}(2016)}]{Hoffmann-2016}%
  \BibitemOpen
  \bibfield  {author} {\bibinfo {author} {\bibfnamefont {D.}~\bibnamefont
  {Hoffmann}},\ }\href
  {https://www.ifo.de/DocDL/sd-2016-13-hoffman-anfa-irland{\%}20-2016-07-14.pdf}
  {\bibinfo {title} {{Erste Erkenntnisse zum ANFA-Abkommen: ANFA erm\"oglicht
  Finanzierung von Bankenabwicklungen durch nationale Zentralbanken}}}
  (\bibinfo {year} {2016}),\ \bibinfo {note} {ifo Schnelldienst 13/2016 - 69.
  Jahrgang - 14. Juli 2016, accessed 5. Sep. 2024}\BibitemShut {NoStop}%
\bibitem [{\citenamefont {{European Central
  Bank}}(2024{\natexlab{b}})}]{EuropeanCentralBank2024Sep}%
  \BibitemOpen
  \bibfield  {author} {\bibinfo {author} {\bibnamefont {{European Central
  Bank}}},\ }\bibfield  {title} {\bibinfo {title} {{What is ANFA?}},\ }\href
  {https://www.ecb.europa.eu/ecb-and-you/explainers/tell-me-more/html/anfa\_qa.en.html}
  {\bibfield  {journal} {\bibinfo  {journal} {European Central Bank}\ }
  (\bibinfo {year} {2024}{\natexlab{b}})},\ \bibinfo {note} {european Central
  Bank, publishing date February 5, 2016, updated on September 13, 2024,
  accessed December 10, 2024}\BibitemShut {NoStop}%
\bibitem [{\citenamefont {{European Central
  Bank}}(2022)}]{EuropeanCentralBank2022Dec}%
  \BibitemOpen
  \bibfield  {author} {\bibinfo {author} {\bibnamefont {{European Central
  Bank}}},\ }\href
  {https://eur-lex.europa.eu/legal-content/EN/TXT/PDF/?uri=IMMC:AGR/2022/12191}
  {\bibinfo {title} {{Agreement on Net Financial Assets of 19 December 2022}}}
  (\bibinfo {year} {2022}),\ \bibinfo {note} {european Central Bank, publishing
  date December 19, 2022, accessed September 8th, 2024}\BibitemShut {NoStop}%
\bibitem [{\citenamefont {Hannoun}\ \emph {et~al.}(2019)\citenamefont
  {Hannoun}, \citenamefont {Issing}, \citenamefont {Liebscher}, \citenamefont
  {Schlesinger}, \citenamefont {Stark}, \citenamefont {Wellink}, \citenamefont
  {de~Larosi\`ere},\ and\ \citenamefont {Noyer}}]{ECBMemorandum2019Oct}%
  \BibitemOpen
  \bibfield  {author} {\bibinfo {author} {\bibfnamefont {H.}~\bibnamefont
  {Hannoun}}, \bibinfo {author} {\bibfnamefont {O.}~\bibnamefont {Issing}},
  \bibinfo {author} {\bibfnamefont {K.}~\bibnamefont {Liebscher}}, \bibinfo
  {author} {\bibfnamefont {H.}~\bibnamefont {Schlesinger}}, \bibinfo {author}
  {\bibfnamefont {J.}~\bibnamefont {Stark}}, \bibinfo {author} {\bibfnamefont
  {N.}~\bibnamefont {Wellink}}, \bibinfo {author} {\bibnamefont
  {de~Larosi\`ere}},\ and\ \bibinfo {author} {\bibfnamefont {C.}~\bibnamefont
  {Noyer}},\ }\href
  {https://www.hanswernersinn.de/dcs/Memorand-ECB-Monetary-Policy-04102019.pdf}
  {\bibinfo {title} {Memorandum on the {ECB}'s monetary policy}} (\bibinfo
  {year} {2019}),\ \bibinfo {note} {published October 4, 2019, accessed 9. Sep.
  2024}\BibitemShut {NoStop}%
\bibitem [{\citenamefont {Congdon}(2022)}]{Congdon2022Mar}%
  \BibitemOpen
  \bibfield  {author} {\bibinfo {author} {\bibfnamefont {T.}~\bibnamefont
  {Congdon}},\ }\bibfield  {title} {\bibinfo {title} {Can the {E}urozone manage
  its free rider problem?},\ }\href
  {https://feu-journal.eu/wp-content/uploads/2023/03/FEU-Journal-Issue-2\_Inflation-rising.pdf}
  {\bibfield  {journal} {\bibinfo  {journal} {Future Europe Journal}\ }\textbf
  {\bibinfo {volume} {2}},\ \bibinfo {pages} {61} (\bibinfo {year}
  {2022})}\BibitemShut {NoStop}%
\bibitem [{\citenamefont {Barth}\ and\ \citenamefont
  {Bennett}(1975)}]{barth-1975}%
  \BibitemOpen
  \bibfield  {author} {\bibinfo {author} {\bibfnamefont {J.~R.}\ \bibnamefont
  {Barth}}\ and\ \bibinfo {author} {\bibfnamefont {J.~T.}\ \bibnamefont
  {Bennett}},\ }\bibfield  {title} {\bibinfo {title} {Cost-push versus
  demand-pull inflation: {S}ome empirical evidence},\ }\href
  {https://doi.org/10.2307/1991632} {\bibfield  {journal} {\bibinfo  {journal}
  {Journal of Money, Credit and Banking}\ }\textbf {\bibinfo {volume} {7}},\
  \bibinfo {pages} {391} (\bibinfo {year} {1975})}\BibitemShut {NoStop}%
\bibitem [{\citenamefont {Werner}(2003)}]{Werner2003May}%
  \BibitemOpen
  \bibfield  {author} {\bibinfo {author} {\bibfnamefont {R.}~\bibnamefont
  {Werner}},\ }\href
  {https://quantumpublishers.com/uk-cart/index.php?route=product/product{\&}product\_id=50}
  {\emph {\bibinfo {title} {Princes of the {Y}en: {J}apan's Central Bankers and
  the Transformation of the Economy}}},\ \bibinfo {edition} {2nd}\ ed.\
  (\bibinfo  {publisher} {Routledge},\ \bibinfo {address} {New York, NY, USA},\
  \bibinfo {year} {2003})\BibitemShut {NoStop}%
\bibitem [{\citenamefont {Werner}(2005)}]{Werner2005Mar}%
  \BibitemOpen
  \bibfield  {author} {\bibinfo {author} {\bibfnamefont {R.}~\bibnamefont
  {Werner}},\ }\href {https://doi.org/10.1057/9780230506077} {\emph {\bibinfo
  {title} {New Paradigm in Macroeconomics: {S}olving the Riddle of {J}apanese
  Macroeconomic Performance}}}\ (\bibinfo  {publisher} {Palgrave Macmillan},\
  \bibinfo {address} {London, UK},\ \bibinfo {year} {2005})\BibitemShut
  {NoStop}%
\end{thebibliography}%

\end{document}